# The Inverse Problem of Quartic Photonics

THOMAS MULKEY,[1] JIMMY DILLIES,[2] MAXIM DURACH[1,*]

[1]*Physics Department*, Georgia Southern University, Math/Physics Bldg., 65 Georgia Avenue, Statesboro, GA 30460
[2]*Mathematics Department*, Georgia Southern University, Math/Physics Bldg., 65 Georgia Avenue, Statesboro, GA 30460
*Corresponding author: mdurach@georgiasouthern.edu

**We propose an approach to engineer quartic metamaterials starting from the desired photonic states. We apply our method to the design of the high-k asymptotics of metamaterials, extreme non-reciprocity and complex bi-anisotropic media.**

The main goal of photonics could be defined as designing material structures that support desired electromagnetic field distributions. Consider a monochromatic optical field with frequency $\omega$ whose electric and magnetic fields are given by vector functions $\boldsymbol{E}(\boldsymbol{r})$ and $\boldsymbol{H}(\boldsymbol{r})$. The possibility of creating such fields in a material is contingent on the condition that the plane waves these fields contain in their Fourier expansions

$$\boldsymbol{E}(\boldsymbol{r}) = \int \boldsymbol{E_k}\, exp[i\boldsymbol{kr}]\, d^3r, \qquad \boldsymbol{H}(\boldsymbol{r}) = \int \boldsymbol{H_k}\, exp[i\boldsymbol{kr}]\, d^3r,$$

are supported by the material. A particular electromagnetic plane-wave degree of freedom exists in a medium if it follows Maxwell's equations in *k*-space

$$\boldsymbol{k} \times \boldsymbol{E_k} = k_0 \boldsymbol{B_k} \text{ and } \boldsymbol{k} \times \boldsymbol{H_k} = -k_0 \boldsymbol{D_k}, \qquad (1)$$

where $k_0 = \omega/c$ and vectors $\boldsymbol{D_k} = \boldsymbol{E_k} + 4\pi \boldsymbol{P_k}$ and $\boldsymbol{B_k} = \boldsymbol{H_k} + 4\pi \boldsymbol{M_k}$ contain information about material response via polarization $\boldsymbol{P_k}$ and magnetization $\boldsymbol{M_k}$ vectors. Conventional materials correspond to dispersive, local, and isotropic material relationships $\boldsymbol{D_k} = \varepsilon(\omega) \boldsymbol{E_k}$ and $\boldsymbol{B} = \mu(\omega) \boldsymbol{H}$ [1]. Here we consider the implications of the most general linear local material relationship that can be expressed as [2-7]:

$$\begin{pmatrix} \boldsymbol{D_k} \\ \boldsymbol{B_k} \end{pmatrix} = \widehat{M} \begin{pmatrix} \boldsymbol{E_k} \\ \boldsymbol{H_k} \end{pmatrix}, \widehat{M} = \begin{pmatrix} \hat{\epsilon} & \hat{X} \\ \hat{Y} & \hat{\mu} \end{pmatrix},$$

where $\hat{\epsilon}, \hat{\mu}, \hat{X}, \hat{Y}$ are 3x3 tensors characterizing dielectric permittivity, magnetic permeability and magnetoelectric coupling correspondingly. This includes 36 effective material parameters, which populate the 6x6 transformation matrix $\widehat{M}$ relating $\boldsymbol{D_k}, \boldsymbol{B_k}$ and $\boldsymbol{E_k}, \boldsymbol{H_k}$ (see numerical example in Fig. 1(a)).

In this Letter, we demonstrate that one needs to specify the *k*-vectors and $\boldsymbol{E_k}, \boldsymbol{H_k}$ amplitudes of 6 arbitrary plane waves (with some limitations specified below) to fully define the required 36 material parameters of the material that will support these waves and, correspondingly, all the other fields possible in the bulk of this material. Here we do not discuss the immediate practical availability with the current technology of the different values of material parameters we obtain in our examples, but provide a recipe to obtain the values of the parameters needed for the desired field distributions to inform and drive the future design of the corresponding metamaterials. The interest in metamaterials with extreme and unconventional properties [8], such as negative refraction, hyperbolic dispersion, optical magnetism, anisotropy, chirality, cloaking, supercoupling, non-reciprocity etc. is growing to feed the technological demands of the industries, marketplaces and security [9-10].

Usually one starts with a set of the effective material parameters for a material or metamaterial at hand and finds the possible electromagnetic fields in this material (the direct problem in Fig. 1). Maxwell's equations (1) can be rewritten as $\widehat{Q}\Gamma = \widehat{M}\Gamma$, with $\Gamma = (E_x, E_y, E_z, H_x, H_y, H_z)$ and

$$\widehat{M} = \begin{pmatrix} \hat{\epsilon} & \hat{X} \\ \hat{Y} & \hat{\mu} \end{pmatrix}, \widehat{Q} = \begin{pmatrix} 0 & -\hat{R} \\ \hat{R} & 0 \end{pmatrix}, \hat{R} = \frac{1}{k_0}\begin{pmatrix} 0 & -k_z & k_y \\ k_z & 0 & -k_x \\ -k_y & k_x & 0 \end{pmatrix}.$$

This system has nontrivial solutions if its determinant of matrix $\widehat{\Delta} = \widehat{M} - \widehat{Q}$ is zero

$$\begin{aligned}\text{Det}(\widehat{\Delta}) &= \text{Det}\left(\hat{\epsilon} - (\hat{X} + \hat{R})\hat{\mu}^{-1}(\hat{Y} - \hat{R})\right) \text{Det}(\hat{\mu}) \\ &= \text{Det}\left(\hat{\mu} - (\hat{Y} - \hat{R})\hat{\epsilon}^{-1}(\hat{X} + \hat{R})\right) \text{Det}(\hat{\epsilon}) = 0. \end{aligned} \qquad (2)$$

The determinant (2) is a multivariate quartic function $f = f(k_x, k_y, k_z)$, i.e. it is a polynomial of degree 4. In other words, the *k*-vectors of the plane waves that satisfy Maxwell's equations belong to a quartic surfaces $f(k_x, k_y, k_z) = 0$ in *k*-space (see example in Fig. 1(b)) [3,11-13]. Each point on such a photonic quartic surface corresponds to a solution of Eq. (1), i.e. to a field vector $\Gamma = (E_x, E_y, E_z, H_x, H_y, H_z)$. The quartic surfaces considered in our study correspond to the following equation

$$\sum_{i+j+l+m=4}[\alpha_{ijlm}k_x^i k_y^j k_z^l k_0^m] = 0 \qquad (3)$$

with 35 coefficients $\alpha_{ijlm}$. In the sum the powers $i, j, l, m$ run from 0 to 4 such that $i + j + l + m = 4$.

Mathematically, there is no complete classification or global picture of quartic surfaces. In the case of singular quartics, the essential properties of surfaces with nodes (or nodal curves) are discussed in Ref [14]. A special case of (mildly singular) quartic K3 surfaces correspond to Kummer varieties. These K3 surfaces have the property that they can be seen as the quotient of two tori by an involution having 16 fixed points. The relation between Kummer K3s and the optics of the studied materials has been discussed in Ref [3]. Smooth quartic surfaces are a classical example of K3 surfaces. Their fundamental group is trivial and their canonical class as well. Quartic K3 surfaces are parametrized by 19 moduli (essentially, coefficients). K3 surfaces are very interesting as their properties combine geometric and arithmetic features [15]. Nevertheless, some families of smooth quartic surfaces, such as ruled quartics, have been studied and classified before [16].

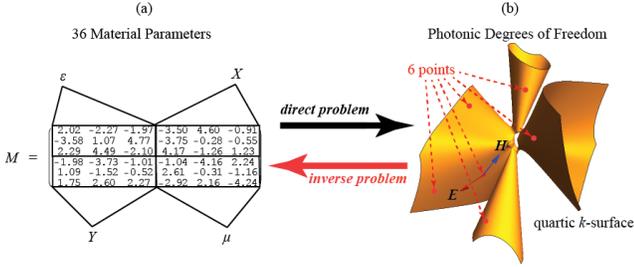

Fig. 1. The direct and inverse problems of quartic photonics. (a) a numerical example of the effective material parameters matrix $\widehat{M}$; (b) the corresponding quartic surface in k-space with 6 points selected for the inverse problem; vector amplitudes of the electric (red) and magnetic (blue) fields are shown for one of the points.

The coefficients $\alpha_{ijlm}$ of the quartic surface Eq. (3) can be explicitly found from the material parameters using Eq. (2). But for the design of the metamaterials, it would be interesting to solve the inverse problem, i.e. to find the effective material parameters of the metamaterials that would support specific plane waves needed for the creation of desired field distributions. The coefficients $\alpha_{ijlm}$ in Eq. (3) are complicated nonlinear functions of the material parameters which cannot be easily inverted to resolve the inverse problem of finding material parameters starting from the desired shape of the quartic surface. In any case, the knowledge of all $\alpha_{ijlm}$ is not the full solution of the direct problem, since it does not contain the information about the field vectors Γ of the supported waves. Therefore, the coefficients $\alpha_{ijlm}$ should not be the input information of the inverse problem.

We formulate the inverse problem (see Fig. 1) as follows. Imagine one needs to create a metamaterial that propagates a set of desired plane waves whose *k*-vectors as well as the field vectors $\Gamma = (E_x, E_y, E_z, H_x, H_y, H_z)$ are given. These plane waves should satisfy Eqs. (1) with the effective material parameters in the unknown matrix $\widehat{M}$. Since we have 36 unknown material parameters, we can specify characteristics of 6 plane waves to form a complete system of equations.

We rewrite Eqs. (1) for the 6 desired waves simultaneously as a matrix equation $\widehat{M}\widehat{G} = \widehat{P}$, where

$$\widehat{G} = \begin{pmatrix} E_{x1} & E_{x2} & E_{x3} & E_{x4} & E_{x5} & E_{x6} \\ E_{y1} & E_{y2} & E_{y3} & E_{y4} & E_{y5} & E_{y6} \\ E_{z1} & E_{z2} & E_{z3} & E_{z4} & E_{z5} & E_{z6} \\ H_{x1} & H_{x2} & H_{x3} & H_{x4} & H_{x5} & H_{x6} \\ H_{y1} & H_{y2} & H_{y3} & H_{y4} & H_{y5} & H_{y6} \\ H_{z1} & H_{z2} & H_{z3} & H_{z4} & H_{z5} & H_{z6} \end{pmatrix}$$

is the matrix whose columns are vectors $\Gamma_i$, while columns of matrix $\widehat{P}$ are vectors $\widehat{Q}_i\Gamma_i$, where $i$ runs from 1 to 6. The antisymmetric matrix $\widehat{Q}_i$ is composed from the *k*-vector components, while $\Gamma_i$ from fields of the 6 desired waves. Now this system can be solved and the effective material parameters matrix can be found explicitly from the parameters of the 6 chosen waves

$$\widehat{M} = \widehat{P}\widehat{G}^{-1} \qquad (4)$$

Finding matrix $\widehat{M}$ with the unknown material parameters gives the solution to the problem of finding a metamaterial that supports a desired field distribution (i.e. the inverse problem in Fig. 1).

Hyperbolic metamaterials, a class of "quadratic" metamaterials (i.e. their k-surfaces are described by quadratic equations), have generated huge interest, since they support high-k photonic states, a property which leads to diverging photonic density of states allowing spontaneous and thermal emission engineering [17-20]. In the quartic metamaterials we study here Eqs. (2)-(3) may have asymptotic solutions with large *k*-vectors $k = \sqrt{k_x^2 + k_y^2 + k_z^2} \gg k_0$ as well. In this asymptotic limit only the 4th order terms with $m = 0$ in Eq. (3) matter.

When considering the asymptotic behaviour of a quartic k-surface, one can intersect the quartic surface with a sphere of radius $k$ and look at the behavior in the limit $k \gg k_0$. Consider a spherical coordinate system in k-space $k_x = k \sin\theta \sin\varphi$, $k_y = k \sin\theta \cos\varphi$, $k_z = k \cos\theta$. In this coordinate system Eq. (3) in the limit $k \gg k_0$ asymptotically reads

$$\sum_{i+j+l=4}[\alpha_{ijl0}(\sin\theta \sin\varphi)^i (\sin\theta \cos\varphi)^j (\cos\theta)^l] = 0 \quad (5)$$

The solution of Eq. (5) is a function $f_{as}(\theta,\varphi)$ which gives directions in *k*-space corresponding to asymptotic solutions of Eqs. (2)-(3). In general, there are 15 parameters $\alpha_{ijl0}$ in Eq. (4) which can be manipulated to obtain different asymptotics $f_{as}(\theta,\varphi)$. For example, the topology of the k-surface of a hyperbolic material in the asymptotic limit corresponds to two circles on a cross-section with a sphere with $k \gg k_0$. The more general class of quartic metamaterials, in principle, provides rich opportunity to engineer these bulk high-k states. To classify the quartic k-surfaces in the asymptotic limit topologically it is convenient to use the projective plane P3. The classification is provided by the following theorem [21]: A *smooth projective real* quartic curve consists *topologically* of: (i) 1 circle; (ii) 2 adjacent circles (example in Fig. 2 or hyperbolic materials); (iii) 2 nested circles; (iv) 3 circles; (v) 4 circles (example in Fig. 1); (vi) an empty set (e.g. vacuum).

We demonstrate how our solution to the inverse problem applies to the engineering of the high-*k* behavior. For illustration, we start with a randomly chosen effective material parameters matrix

$$\widehat{M} = \begin{pmatrix} -3.15 & 4.81 & 4.47 & -1.66 & -0.52 & -2.18 \\ -2.82 & 2.81 & 0.17 & 2.74 & -3.71 & 1.06 \\ -1.55 & 1.60 & 0.38 & -0.56 & -4.37 & -0.66 \\ -2.29 & 4.24 & -2.95 & -2.23 & 3.68 & -4.43 \\ -3.30 & 4.22 & -2.06 & -0.98 & 4.51 & 1.00 \\ 2.96 & -4.58 & 1.23 & -3.31 & -2.70 & 4.19 \end{pmatrix}. \quad (6)$$

Solving Eqs. (1)-(2) we arrive at the quartic surface shown in Fig. 2. In Fig. 2(a) we show a cross-section of this surface at $k = 100\,k_0$ (red curve) and the corresponding asymptotic solution $f_{as}(\theta,\varphi)$ of Eq. (5) (black). From this high-k cross-section, 6 points are randomly selected, shown in green in both panels Fig. 2(a) and (b). In Fig. 2(b), the quartic surface is plotted in Cartesian coordinates with the selected 6 points and the corresponding field amplitudes shown by arrows (electric fields are red, magnetic – blue). Retaining the data of only the selected 6 points, we solve Eq. (4) and get the parameters of matrix (6) back.

The field vectors in Fig. 2(b) show a major property of the high-k quasistatic electromagnetic fields. Both electric and magnetic fields are directed with high precision in the radial direction, i.e. they are parallel to the k-vectors $E, H \parallel k$. Since in our inverse problem approach the fields can be chosen arbitrarily, it is interesting to see what would happen if we change the direction of the fields of one of the waves from longitudinal to, e.g., transverse, i.e. break the quasistatic $E, H \parallel k$ relationship. In this case, the elements of the resulting material matrix become larger than in Eq. (6), which leads to a shift of the applicability of the asymptotic approximation of Eq. (5) to higher $k$.

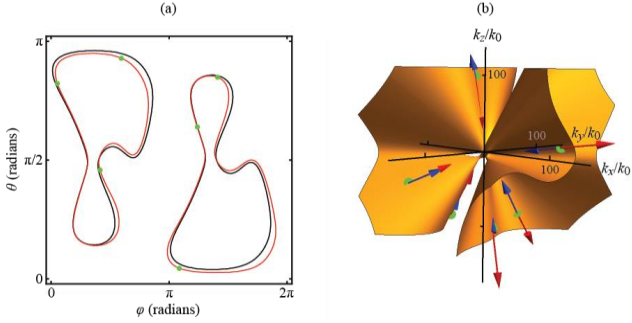

Fig. 2. Numerical example of asymptotic behavior. (a) asymptotic k-surface in $(\varphi, \theta)$ coordinates (black) compared to $k = 100k_0$ cross-section (red); green points are selected at this cross-section for effective material parameters retreaval; (b) the corresponding quartic surface in k-space with the same green points indicated with vector amplitudes of electric (red) and magnetic (blue) fields. Note that the fields are practically parallel to the k vectors as they should in the high-k quasistatic limit.

Now let us turn to the fundamental limitations on the possible electromagnetic fields in the quartic metamaterials that follow directly from our approach. In isotropic materials, i.e. for $\hat{\epsilon} = \epsilon \cdot \hat{1}, \hat{X} = X \cdot \hat{1}, \hat{Y} = Y \cdot \hat{1}$, and $\hat{\mu} = \mu \cdot \hat{1}$, the k-surfaces are quadratic and correspond to a pair of spheres

$$k^4 + k^2 k_0^2 (X^2 + Y^2 - 2\epsilon\mu) + k_0^4 (XY - \epsilon\mu)^2 = 0. \quad \textbf{(7)}$$

The radii of the spheres are given by

$$k^2 = k_0^2 \left( \epsilon\mu - \frac{X^2 + Y^2}{2} \pm \frac{X - Y}{2} \sqrt{(X + Y)^2 - 4\epsilon\mu} \right). \quad \textbf{(8)}$$

and coincide if $X = Y$. If one selects a direction $(\theta, \varphi)$ in k-space one will find 2 waves on the surface Eq. (7)-(8) propagating in this direction with 2 different polarizations and 2 waves in the opposite direction (see Fig. 3(a)).

In a general quartic material such separation of polarizations and quadratic property disappears. In any given direction $(\theta, \varphi)$ Eqs. (2)-(3) take the form

$$\sum_{i+j+l+m=4} \left[ \alpha_{ijlm} (\sin\theta \sin\varphi)^i (\sin\theta \cos\varphi)^j (\cos\theta)^l \left(\frac{k}{k_0}\right)^{4-m} \right] = 0. \quad \textbf{(9)}$$

Eq. (9) is a single-variable quartic equation with respect to $k/k_0$, which in general has up to 4 roots representing 4 different plane waves propagating in the direction $(\theta, \varphi)$. From this a fundamental limitation follows that one cannot require all 6 wave in the inverse problem to have parallel k-vectors.

Nevertheless, we can select any 4 values of the k-number in any direction and 2 more in other directions. As an example, we were able to obtain a quartic material (Fig. 4(b), "reindeer Rudolph" k-surface) with the following effective parameters matrix:

$$\hat{M} = \begin{pmatrix} -4.34 & 3.11 & 10.63 & 5.57 & 3.41 & -10.10 \\ -0.50 & 0.04 & -3.81 & -1.24 & 1.98 & 1.56 \\ 3.45 & -2.77 & -12.72 & 1.27 & 0.93 & -1.16 \\ 1.97 & -5.38 & 2.63 & -5.12 & -2.14 & -0.55 \\ 4.48 & -3.82 & -3.78 & -2.42 & -0.90 & 0.19 \\ -0.26 & -0.79 & 0.89 & -1.18 & -0.15 & 0.90 \end{pmatrix}. \quad \textbf{(10)}$$

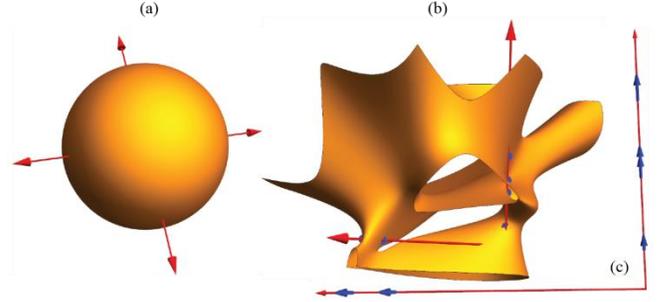

Fig. 3. k-surface engineering; (a) k-surface of an isotropic material; a line in any direction intersects 2 states; (b)-(c) "reindeer Rudolph" k-surface features extremely non-reciprocal behavior, with k-vectors (b) and Poynting vectors (c) of 4 states pointing in the same direction.

In this material all 4 plane waves in the direction $(\theta, \varphi) = (0.11\pi, 1.71\pi)$ have wave-vectors which point in the same direction and there are no waves with opposite k-vectors (Fig. 3(b)). Note that directions of the k-vectors correspond to the direction of the phase propagation. We were able to select the field vectors $\Gamma_i$ such that the Poynting vectors for the 6 waves point in the same directions as k-vectors [Poynting vectors for the waves in Fig 3(b) are shown in Fig. 3(c)]. This means we were able to design all 4 waves in a direction $(\theta, \varphi)$ to propagate both energy and phase in this direction - an example of extreme non-reciprocity [2,22-25].

In some cases the metamaterials are subject to restrictions, e.g. if some of the effective material parameters are known *a priori*. In such situations one could require existence of less than 6 waves and complement the truncated Eqs. (6) with extra restrictions on material parameters. For example, one could require existence of 3 specific waves in a metamaterial and matrices $\hat{G}$ and $\hat{P}$ in Eq. (4) would have 3 columns instead of 6. Then Eq. (4) are rewritten as

$$\hat{\epsilon} = -\hat{E}^{-1}\hat{P}_H - \hat{E}^{-1}\hat{X}\hat{H}, \quad \hat{\mu} = \hat{H}^{-1}\hat{P}_E - \hat{H}^{-1}\hat{Y}\hat{E}, \quad \textbf{(11)}$$

i.e. the permittivity and permeability matrices $\hat{\epsilon}$ and $\hat{\mu}$ can be expressed via the matrices $\hat{X}$ and $\hat{Y}$, k-vectors of the selected waves and the amplitudes of their fields in the matrices

$$k_0 \hat{P}_E = (\mathbf{k_1} \times \mathbf{E_1}, \mathbf{k_2} \times \mathbf{E_2}, \mathbf{k_3} \times \mathbf{E_3})$$
$$k_0 \hat{P}_H = (\mathbf{k_1} \times \mathbf{H_1}, \mathbf{k_2} \times \mathbf{H_2}, \mathbf{k_3} \times \mathbf{H_3})$$
$$\hat{E} = \begin{pmatrix} E_{x1} & E_{x2} & E_{x3} \\ E_{y1} & E_{y2} & E_{y3} \\ E_{z1} & E_{z2} & E_{z3} \end{pmatrix} \text{ and } \hat{H} = \begin{pmatrix} H_{x1} & H_{x2} & H_{x3} \\ H_{y1} & H_{y2} & H_{y3} \\ H_{z1} & H_{z2} & H_{z3} \end{pmatrix}.$$

The inspection of the modified inverse solution of the photonic problem given by Eqs. (11) leads to the conclusion that if the magnetoelectric coupling $\hat{X}$ and $\hat{Y}$ is fixed one can freely select only 3 plane waves. Eqs (11) are particularly convenient for the design of materials in absence of magnetoelectric coupling $\hat{X} = \hat{Y} = \hat{0}$ in which case $\hat{\epsilon} = -\hat{E}^{-1}\hat{P}_H$, $\hat{\mu} = \hat{H}^{-1}\hat{P}_E$.

In another example we apply Eqs. (11) as shown in Fig. 4 where the magnetoelectric coupling is selected in the chiral form $\hat{X} = -\hat{Y} = -i\hat{1}$ and the medium supports 3 of the 6 waves selected for Fig. 3(b-c). This is an example of a general situation, when matrix $\hat{M}$ is complex

$$\text{Re}[\widehat{M}] = \begin{pmatrix} 18.6 & -15.4 & -18.2 & 0 & 0 & 0 \\ 2.41 & -1.48 & 7.94 & 0 & 0 & 0 \\ 7.21 & -6.01 & -16.3 & 0 & 0 & 0 \\ 0 & 0 & 0 & 2.34 & 2.59 & 3.73 \\ 0 & 0 & 0 & -1.00 & 0.24 & -1.13 \\ 0 & 0 & 0 & 0.62 & 0.86 & 2.18 \end{pmatrix}. \quad \textbf{(12a)}$$

$$\text{Im}[\widehat{M}] = \begin{pmatrix} -0.39 & -0.23 & -2.56 & -1 & 0 & 0 \\ 2.52 & -1.78 & 2.52 & 0 & -1 & 0 \\ -1.64 & 1.11 & 2.29 & 0 & 0 & -1 \\ 1 & 0 & 0 & 1.31 & 0.44 & 0.97 \\ 0 & 1 & 0 & 1.88 & 0.97 & 1.03 \\ 0 & 0 & 1 & 0.02 & -0.15 & -0.24 \end{pmatrix} \quad \textbf{(12b)}$$

If matrix $\widehat{M}$ is complex the coefficients $\alpha_{ijlm}$ in Eq. (3) are complex too and Eqs. (2)-(3) correspond to two quartic surfaces with coefficients $\text{Re}[\alpha_{ijlm}]$ and $\text{Im}[\alpha_{ijlm}]$ (orange and blue in Fig. 4(a)). The solutions of Eqs. (2)-(3) with real $k_0$ and $\boldsymbol{k}$ are at the intersection of these quartic surfaces which generally is a curve in k-space (blue in Fig. 4(b)). One can see how the solution curve passes through the desired 3 waves (3 of the 6 waves highlighted in green in Fig. 4(b) and blue in Fig. 3(b)). Note that the solution vectors $\Gamma_i$ for the selected 3 waves are the same as in the case of the "reindeer Rudolph" k-surface in Fig. 3(b-c), i.e. they are real and represent linear polarization, in contrast with the isotropic chiral materials (see Eqs. (7-8) for $X = -Y = -i$) in which the solution vectors are complex, representing circular polarization.

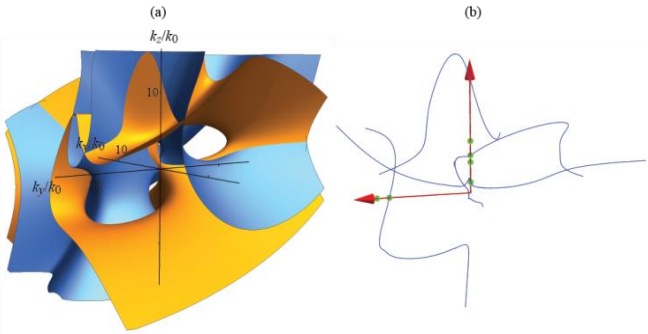

Fig. 4. k-surface with complex material parameters. (a) Two quartic surfaces with coefficients $\text{Re}[\alpha_{ijlm}]$ (orange) and $\text{Im}[\alpha_{ijlm}]$ (blue); (b) the intersection of the quartic surfaces of panel (a) shown as blue curve of solutions of Eqs. (2)-(3). The desired 3 waves are highlighted in green and belong to the curve.

To summarize, we have formulated and solved an inverse problem of photonics which enables one to find effective material parameters of a quartic or quadratic metamaterial that supports a set of desired plane waves with customizable k-vectors and field amplitudes. We demonstrate the power of our approach by studying the high-k asymptotics of quartic metamaterials, extreme non-reciprocity design and metamaterials with complex bi-anisotropic parameters.

**Funding.** College Office of Undergraduate Research (COUR) Research Award, Georgia Southern University.

## References

1. L. D. Landau, E. M. Lifshitz and L. P. Pitaevskii, Electrodynamics of continuous media. Vol. 8, (Elsevier, 2013).
2. A. Sihvola, Metamaterials in electromagnetics, Metamaterials 1, pp. 2–11, (2007).
3. P. Baekler, A. Favaro, Y. Itin, F.W. Hehl, The Kummer tensor density in electrodynamics and in gravity. Annals of Physics, 349, pp.297-324 (2014).
4. F. W. Hehl, Yu. N. Obukhov, J. -P. Rivera and H. Schmid, Relativistic nature of a magnetoelectric modulus of Cr2O3 crystals: A fourdimensional pseudoscalar and its measurement, Phys. Rev. A 77, p. 022106 (2008)
5. I.V. Lindell, A.H. Sihvola, S.A. Tretyakov, A.J. Viitanen, Electromagnetic Waves in Chiral and Bi-isotropic Media (Artech House, Norwood, MA, 1994).
6. A. Serdyukov, I. Semchenko, S. Tretyakov, A. Sihvola, Electromagnetics of Bi-anisotropic Materials: Theory and Applications. (Gordon and Breach, Amsterdam, 2001).
7. F. W. Hehl and Yu. N. Obukhov, Foundations of Classical Electrodynamics: Charge, flux, and metric (Birkhauser, Boston, 2003)
8. Sihvola, A., S. Tretyakov, and A. De Baas. "Metamaterials with extreme material parameters." Journal of Communications Technology and Electronics 52(9), pp. 986-990 (2007).
9. M. A. Noginov, and V. A. Podolskiy, eds. Tutorials in metamaterials (CRC press, 2011).
10. M. A, Noginov, G. Dewar, M. W. McCall, and N. I. Zheludev, Tutorials in complex photonic media, SPIE, 2009.
11. I.V. Lindell, Plane-Wave Propagation in Electromagnetic PQ Medium, Progress In Electromagnetics Research, Vol. 154, pp.23-33 (2015).
12. K. A. Vytovtov, "Investigation of the plane wave behavior within bianisotropic media." In Mathematical Methods in Electromagnetic Theory, Vol. 1, pp. 326-328 (2000).
13. A. Favaro, "Electromagnetic wave propagation in metamaterials: a visual guide to Fresnel-Kummer surfaces and their singular points." In 2016 URSI International Symposium on Electromagnetic Theory (EMTS), pp. 622-624, IEEE, 2016
14. Jessop, C. Quartic Surfaces with Singular Points. Cambridge University Press, 1916
15. D. Huybrechts, Lectures on K3 surfaces, Vol. 158. Cambridge University Press, 2016.
16. K. Rohn, "Die verschiedenen Arten der Regelflächen vierter Ordnung", (1886). Mathematische Abhandlungen aus dem Verlage Mathematischer Modelle von Martin Schilling. Halle a. S., 1904. Mathematische Annalen, vol. 28, No. 2, pp. 284–308 (1886)
17. Z. Jacob, I. Smolyaninov, E. Narimanov. "Broadband Purcell effect: Radiative decay engineering with metamaterials." Applied Physics Letters 100 (18), p. 181105 (2012)
18. M. A. Noginov, H. Li, Yu A. Barnakov, D. Dryden, G. Nataraj, G. Zhu, C. E. Bonner, M. Mayy, Z. Jacob, E. E. Narimanov. "Controlling spontaneous emission with metamaterials." Optics letters 35 (11), pp. 1863-1865 (2010).
19. Y. Guo, C. Cortes, S. Molesky, Z. Jacob. "Broadband super-Planckian thermal emission from hyperbolic metamaterials." Applied Physics Letters 101 (13), p. 131106 (2012).
20. I. Iorsh, A. Poddubny, A. Orlov, P. Belov, Y. S. Kivshar. "Spontaneous emission enhancement in metal–dielectric metamaterials." Physics Letters A 376 (3), pp. 185-187 (2012)`.
21. H.G. Zeuthen, Sur les différentes formes des courbes planes du quatrième ordre, Mathematische Annalen 7, pp. 408–432 (1873)
22. B. D. H. Tellegen, Philips Res. Rep. 3, p. 81 (1948)
23. Tretyakov, S. A., A. H. Sihvola, A. A. Sochava, and C. R. Simovski. "Magnetoelectric interactions in bi-anisotropic media." Journal of Electromagnetic Waves and Applications 12(4), p. 481-497 (1998)
24. C. Rüter, K. G. Makris, R. El-Ganainy, D. N. Christodoulides, M. Segev, and D. Kip. "Observation of parity–time symmetry in optics." Nature Physics 6 (3), pp. 192-195 (2010)
25. Bliokh, Konstantin Y., Yuri S. Kivshar, and Franco Nori. "Magnetoelectric effects in local light-matter interactions.", Physical review letters 113 (3), pp. 033601 (2014)